\documentclass[prl,aps,twocolumn,superscriptaddress]{revtex4-1}
\usepackage{graphicx,color}
\usepackage{amsthm}
\usepackage{amsfonts}
\usepackage{algorithmic}
\usepackage{enumerate}
\usepackage{latexsym}
\usepackage{amsmath}
\usepackage{amssymb}
\usepackage[colorlinks=true,citecolor=blue,linkcolor=blue]{hyperref}

\emergencystretch=\maxdimen
\begin{document}

\title{Quantum Monte Carlo study of superconductivity in rhombohedral trilayer graphene under an electric field}

\author{Huijia Dai}
\affiliation{Department of Physics, Beijing Normal University, Beijing 100875, China\\}
\author{Runyu Ma}
\affiliation{Department of Physics, Beijing Normal University, Beijing 100875, China\\}
\author{Xiao Zhang}
\affiliation{Department of Physics, Beijing Normal University, Beijing 100875, China\\}
\author{Ting Guo}
\affiliation{Department of Physics, Beijing Normal University, Beijing 100875, China\\}

\author{Tianxing Ma}
\email{txma@bnu.edu.cn}
\affiliation{Department of Physics, Beijing Normal University, Beijing 100875, China\\}
\affiliation{Key Laboratory of Multiscale Spin Physics(Ministry of Education), Beijing Normal University, Beijing 100875, China\\}

\begin{abstract}
By using the constrained-phase quantum Monte Carlo method, we performed a systematic study of the ground state of the half filled Hubbard model for a trilayer honeycomb lattice.
We analyze the effect of the perpendicular electric field on the electronic structure, magnetic property and pairing correlations.
It is found that the antiferromagnetism is suppressed by the perpendicular electric field, especially the long-range
parts, and the dominant magnetic fluctuations are still antiferromagnetic. The electronic correlation drives a
$d+id$ superconducting pairing to be dominant over other pairing patterns among various electric fields and interaction
strengths. We also found that the $d+id$ pairing correlation is greatly enhanced as the on-site Coulomb interaction is increased.
Our intensive numerical results may unveil the nature of the recently observed superconductivity in rhombohedral trilayer graphene under an electric field.
\end{abstract}

\noindent


\maketitle

\begin{figure}[tbp]
\includegraphics[scale=0.45]{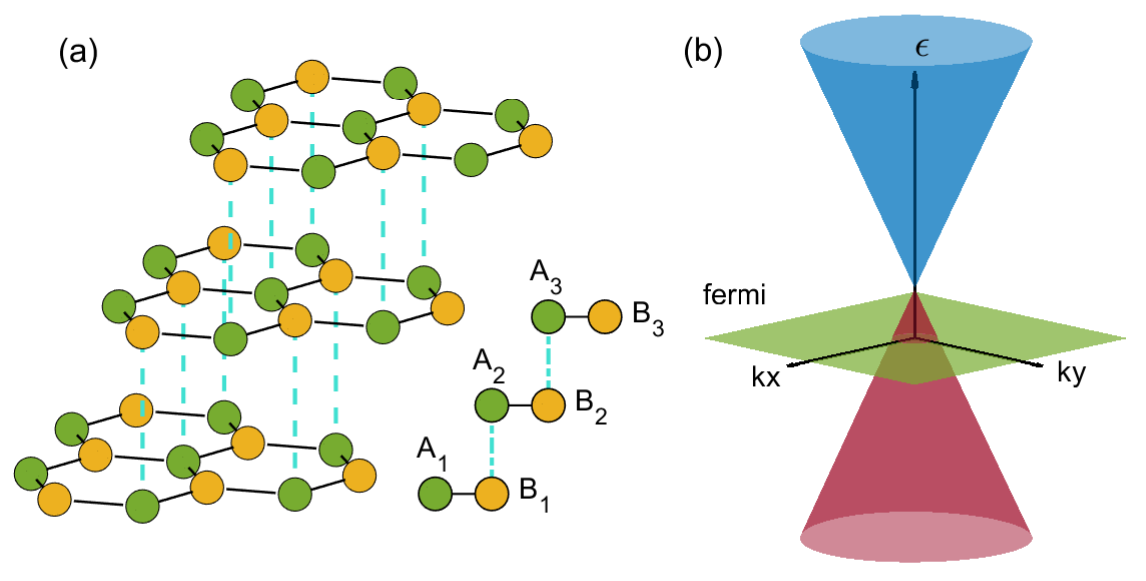}
\caption{(a) Crystal structure diagram for rhombohedral trilayer graphene (ABC-TLG).
Green (yellow) dots represent sublattice A (B) with the subscript denoting the layer index $i = 1, 2, 3$ of the sites.
Each $A_2$ and $B_1$ site, as well as each $A_3$ and $B_2$ site, overlap.
(b) Band structure around the $K$ point of the honeycomb lattice.
The induced perpendicular electric field leads to a shift in the Fermi surface.
} \label{Structure}
\end{figure}

\noindent
{\it Introduction}.
The experimental discovery of superconductivity and correlated insulating states in magic-angle twisted bilayer graphene (TBG) \cite{Cao2018A,Cao2018B,Yankowitz1059} has led to rapid development of research focused on trilayer graphene (TLG) and multilayer graphene systems.
Typically, there are three possible arrangements of graphene layers: AAA stacking, ABA stacking, and ABC stacking corresponding to hexagonal, Bernal, and rhombohedral graphene, respectively.
As interlayer coupling strongly modifies the linear dispersion of monolayer graphene, the electronic structures vary in multilayer graphene films.
The unique electronic structure of multilayer graphene largely raises the possibility of serving as a new platform for unknown physics, and substantial experimental efforts have gone into this field.
It has been reported that a gate-tunable Mott insulator and signatures of superconductivity are observed in a rhombohedral trilayer graphene (ABC-TLG) heterostructures with a moir\'{e} superlattice \cite{NaturePhysics_Chen,Nature572_Chen,Nature579_Chen}, and the crystal structure diagram for ABC-TLG is illustrated in Fig. \ref{Structure}(a).
Displacement field--tunable superconductivity is also discovered in alternating-twist magic-angle trilayer graphene \cite{science.abg0399}.
In twisted double bilayer graphene, Shen \emph{et al}. reported the discovery and characterization of displacement field-tunable electronic phases \cite{Shen2020}.
In twisted bilayer-bilayer graphene, Cao \emph{et al}. found a rich phase diagram with tunable correlated insulator states and spin-polarized phases \cite{Cao2020}.
Culmination of these recent experimental achievements has motivated theoretical studies on exotic correlated electronic phases in graphene superlattices \cite{PhysRevB.99.205150,PhysRevLett.122.016401,Lee2019,PhysRevLett.123.197702,PhysRevLett.121.087001,PhysRevB.99.075127,PhysRevLett.124.187601,Pantale_n_2021}.

The application of a perpendicular electric field is a common method to change the band structure of graphene systems.
Charge carriers, electrons, or holes can be introduced into graphene through the perpendicular electric field, which generates an interlayer potential asymmetry and can induce an energy gap in the electronic spectrum, modifying the electronic structure near the K point, as illustrated in Fig. \ref{Structure}(b).
Theoretical and experimental studies have shown that the band structure is tunable with the perpendicular electric field in bilayer graphene \cite{PhysRevLett.99.216802,APL92_24,2009Direct}, ABC-TLG \cite{Lui2011,PhysRevB.82.035409,APL_Wu} and multilayer graphene \cite{PhysRevB.79.035421,PhysRevB.80.195401,PhysRevB.81.125304,JPCC_Tang}, which offers an exciting opportunity to investigate the richer electronic structure and widen the range of application for graphene systems in electronics.
More recently, experiments have reported the observation of superconductivity and tunable magnetism in ABC-TLG with an external electric displacement field in the stacking direction \cite{Nature434_Zhou,Nature429_Zhou}.
The electric field changes the cubic band structure of pristine ABC-TLG, which exhibits a rather flat dispersion at low energy and induces layer polarization of the electronic density, leading to a uniform and isotropic gap between the valence and conduction bands.
The observation of superconductivity hosting gate-tuned magnetism in a clean system without a moir\'e pattern can provide a new perspective on the origin of superconductivity in graphene-based systems.

Most strikingly, two distinct superconducting phases, SC1 and SC2, have been discovered in different regions of the phase diagram \cite{Nature434_Zhou}.
The SC1 phase emerges from a paramagnetic normal state and respects the Pauli limit, which implies s-wave spin-singlet superconductivity \cite{PhysRevLett.127.187001,PhysRevLett.127.247001,PhysRevB.105.L081407}.
The SC2 phase occurs within a fully spin-polarized, valley-unpolarized half metal and is insensitive to an applied in-plane Zeeman field.
An acoustic-phonon-mediated superconducting mechanism \cite{PhysRevLett.127.187001} and electron-electron interaction-driven superconducting mechanism \cite{PhysRevLett.127.247001,dong2021superconductivity} have been proposed to explain the observed SC2 phase, but there are still puzzles, and the pair symmetry of the SC2 phase is still under very active debate \cite{PhysRevLett.127.187001,PhysRevLett.127.247001,PhysRevB.105.L081407,PhysRevB.105.075432,chatterjee2021inter,dong2021superconductivity,you2022,PhysRevLett.130.146001}.
In this paper, we make efforts to identify the nature of the observed SC2 superconducting state and the superconductivity we mention in the following refers to the SC2 phase specifically.

Compared to the determinant quantum Monte Carlo (DQMC) method \cite{PhysRevB.41.9301}, the constrained-path quantum Monte Carlo (CPMC) method is believed to be the more appropriate approach where the sign problem is avoided by the constrained-path approximation.
We focus on the superconducting pairing correlation and magnetic correlation in the ground state of ABC-TLG.
Our simulation shows that the system exhibits a short-range antiferromagnetic correlation at half filling and the superconducting pairing correlation with the $d + id$ wave dominates over other pairing symmetries with the perpendicular electric field.
For further study, we also considered the effect of the on-site Coulomb interaction $U$ and it is found that the superconducting pairing correlation with $d + id$ wave symmetry is enhanced by the existence of $U$.
Our study provides a starting point for further theoretical and experimental investigations of correlation effects and superconductivity in ABC-TLG.

\noindent
{\it Model and methods}.
We study the half filled Hubbard model with an electric field on the ABC-TLG lattice, which is sketched in Fig. \ref{Structure}(a).
The corresponding microscopic model is written as follows:

\begin{equation}
\begin{split}
H = &-t\sum_{\langle{ij}\rangle\sigma}\sum_{l=1}^{3}[a_{il\sigma}^{\dagger}b_{jl\sigma}+\text{H}.\text{c}.]\\
&-t_{\bot}\sum_{i\sigma}[b_{i1\sigma}^{\dagger}a_{i2\sigma}+b_{i2\sigma}^{\dagger}a_{i3\sigma}+\text{H}.\text{c}.]\\
&+\frac{\varepsilon}{2}\sum_{i\sigma}(a_{i2\sigma}^{\dagger}a_{i2\sigma}+b_{i2\sigma}^{\dagger}b_{i2\sigma})\\
&+\varepsilon\sum_{i\sigma}(a_{i3\sigma}^{\dagger}a_{i3\sigma}+b_{i3\sigma}^{\dagger}b_{i3\sigma})\\
&+U\sum_{i}\sum_{l=1}^{3}(n_{ilA\uparrow}n_{ilA\downarrow}+n_{ilB\uparrow}n_{ilB\downarrow}),
\end{split}
\label{1}
\end{equation}
where $a_{il\sigma}^{\dagger}(a_{il\sigma})$ creates (annihilates) electrons at site ${\textbf{R}}_{li}^a$ of the $l$ ($l=1,2,3$) layer with spin $\sigma(\sigma = \uparrow ,\downarrow)$ on sublattice A, as well as $b_{il\sigma}^{\dagger}(b_{il\sigma})$ acting on electrons of sublattice B.
Occupy number operators $n_{ilA\sigma} = a_{il\sigma}^{\dagger}a_{il\sigma}$ and $n_{ilB\sigma} = b_{il\sigma}^{\dagger}b_{il\sigma}$.
$t$ denotes the in-plane hopping amplitude between nearest-neighbor (NN) and $t_{\bot}$ is the interlayer hopping amplitude in the direction perpendicular to the NN bond.
We set $t = 1$ as the default energy scale.
Previous calculations and experiments \cite{RevModPhys.81.109,ROZHKOV20161,PhysRevB.46.4531,Lui2011} indicate that $t = 2.7 \sim 3$ eV and $t_{\bot} = 0.3 \sim 0.4$ eV.
Accordingly, $t_{\bot}=0.1t$ is used in later simulations.
$U$ is the on-site Hubbard repulsive interaction, and $\varepsilon$ is the potential difference, denoting the effect of the applied perpendicular electric field.
The current experimental work on trilayer graphene shows that one can tune the potential difference up to the eV scale \cite{NaturePhysics_Chen,Nature572_Chen,Nature434_Zhou,Nature429_Zhou,lee2019gate}.
For this reason, we selected $0.1t \to 1.0t$ as the parameter range for $\varepsilon$.

Our simulations are mostly performed on the lattice of $L=4$ with periodic boundary conditions.
$L$ is the linear dimension of the lattice.
The number of lattice sites in each layer is $2\times3L^2$, where the number 2 means two inequivalent triangular sublattices and the number 3 means that each triangular sublattice is consistent of 3 rhombus lattices with $L^2$ sites;
the rhombus lattice approximately describes the unit cell as shown in Fig.  \ref{Unit cell}.
The total number is equal to $N_S = 3\times2\times3L^2$.
For the case of half filling, the total electron number is also equal to $3\times2\times3L^2$.

We adopt the constrained-path quantum Monte Carlo (CPMC) method \cite{PhysRevLett.74.3652,PhysRevB.55.7464,PhysRevLett.78.4486,PhysRevB.63.115112,PhysRevB.64.205101,PhysRevB.84.121410} to study the pairing symmetry and magnetic properties.
In the CPMC method, the ground-state wave function $|\phi\rangle$ is projected from an initial wave function $|\Psi_0\rangle$ by a branching random walk in an overcomplete space of constrained Slater determinants $|\Psi_0\rangle$.
The constrained Slater determinant spaces have positive overlaps, where the trial wave function is already known.
Therefore, we can write $|\Psi_0\rangle = \sum\phi\chi(\phi)|\phi\rangle$ in such a space, where $\chi(\phi) > 0$.
After the random walk, we obtain an ensemble of $\phi$, named random walkers.
Thus $|\Psi_0\rangle$ is distributed in the sense of Monte Carlo sampling of $\chi(\phi)$.
A constrained-path approximation is adopted to prevent the sign problem.
In this work, we focused on closed-shell cases and the corresponding free-electron wave functions were chosen to be $|\Psi_T\rangle$.
In a typical CPMC run, the average number of random walkers is set to be $600$ and the time step $\Delta\tau = 0.05$; $40$ blocks of $320$ Monte Carlo steps are used to ensure statistical independence.
After the simulations reach equilibrium, the expectation values for some physical observable $\mathcal{O}$ are estimated based on the backpropagation (BP) technique \cite{PhysRevB.55.7464}.
A comparison of the results obtained from different BP steps indicates that $40$ BP steps can ensure convergence in the regime $U < 4.0t$.

\begin{figure}[tbp]
\includegraphics[scale=0.4]{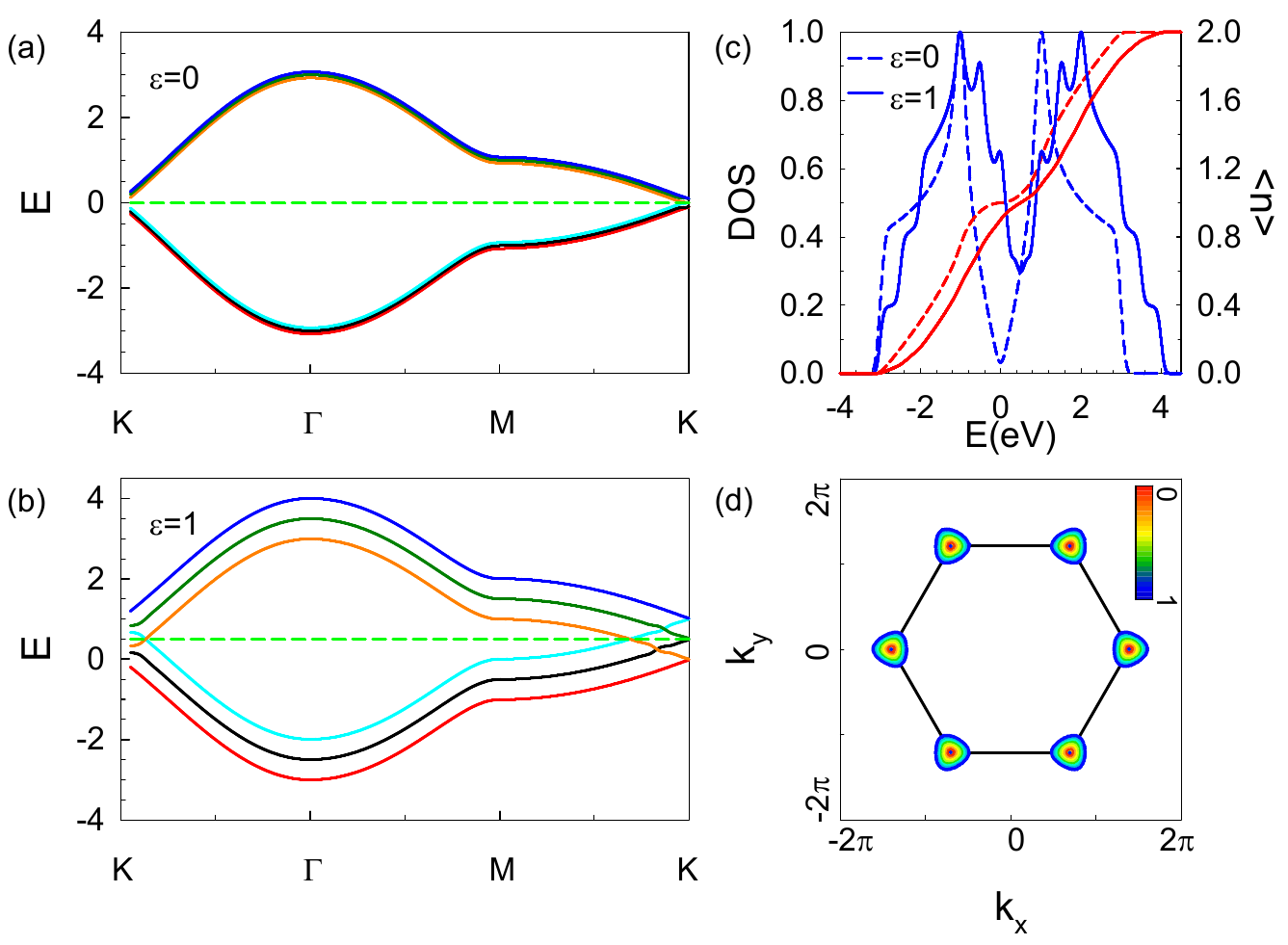}
\caption{(a), (b) Band structures, (c) density of states and (d) Fermi surfaces of the noninteracting Hubbard model on the ABC-TLG for different potential differences $\varepsilon$.
The coordinates of $\Gamma$, $M$, and $K$ shown in (a) and (b) are $(0,0)$, $(0,\frac{2\pi}{\sqrt{3}})$, and $(\frac{2\pi}{3},\frac{2\pi}{\sqrt{3}})$.
The green dashed lines represent the positions of the Fermi levels.
The blue and red curves shown in (c) indicate the density of states and filling density and the solid and dashed lines represent $\varepsilon =0.0$ and $\varepsilon = 1.0$, respectively.
The color bar shown in (d) represents $\varepsilon$ over a range of $0 < \varepsilon <1$.
} \label{Band structure}
\end{figure}

\noindent
{\it Results and discussion}.
We first examine the impact of the electric field on the band structure, Fermi surface and density of states (DOS). 
At $\varepsilon = 0$, as shown in Fig. \ref{Band structure}(a), the three valence bands marked by red, black and cyan lines are almost degenerate and the three conduction bands exhibit similar behavior, which are marked by orange, green and blue lines.
At $\varepsilon = 1$, the originally nearly degenerate bands split, as shown in Fig. \ref{Band structure}(b), and the Fermi level shifts upwards by approximately $0.5 eV$.
Figure \ref{Band structure}(c) clearly shows that the DOS near the Fermi level is dramatically increased and that the van Hove singularity (VHS) is split as $\varepsilon$ increases from $0$ to $1$.
As shown in Fig. \ref{Band structure}(d), it is clear that the Fermi surface gradually expands from a point to a large ring surrounding the $K$ point with increasing $\varepsilon$.
Superconductivity has been found in regimes where the normal-state Fermi surface (FS) has an annular shape; an annular FS is beneficial for an electronic mechanism for superconductivity driven by repulsive Coulomb interactions \cite{PhysRevLett.15.524,Kagan_2015,PhysRevB.81.224505,PhysRevB.83.094518,PhysRevB.96.174514}.
The phenomenon of the increasing ring around the $K$ point in FS indicates that the electric field favors superconductivity.

\begin{figure}[tbp]
\includegraphics[scale=0.4]{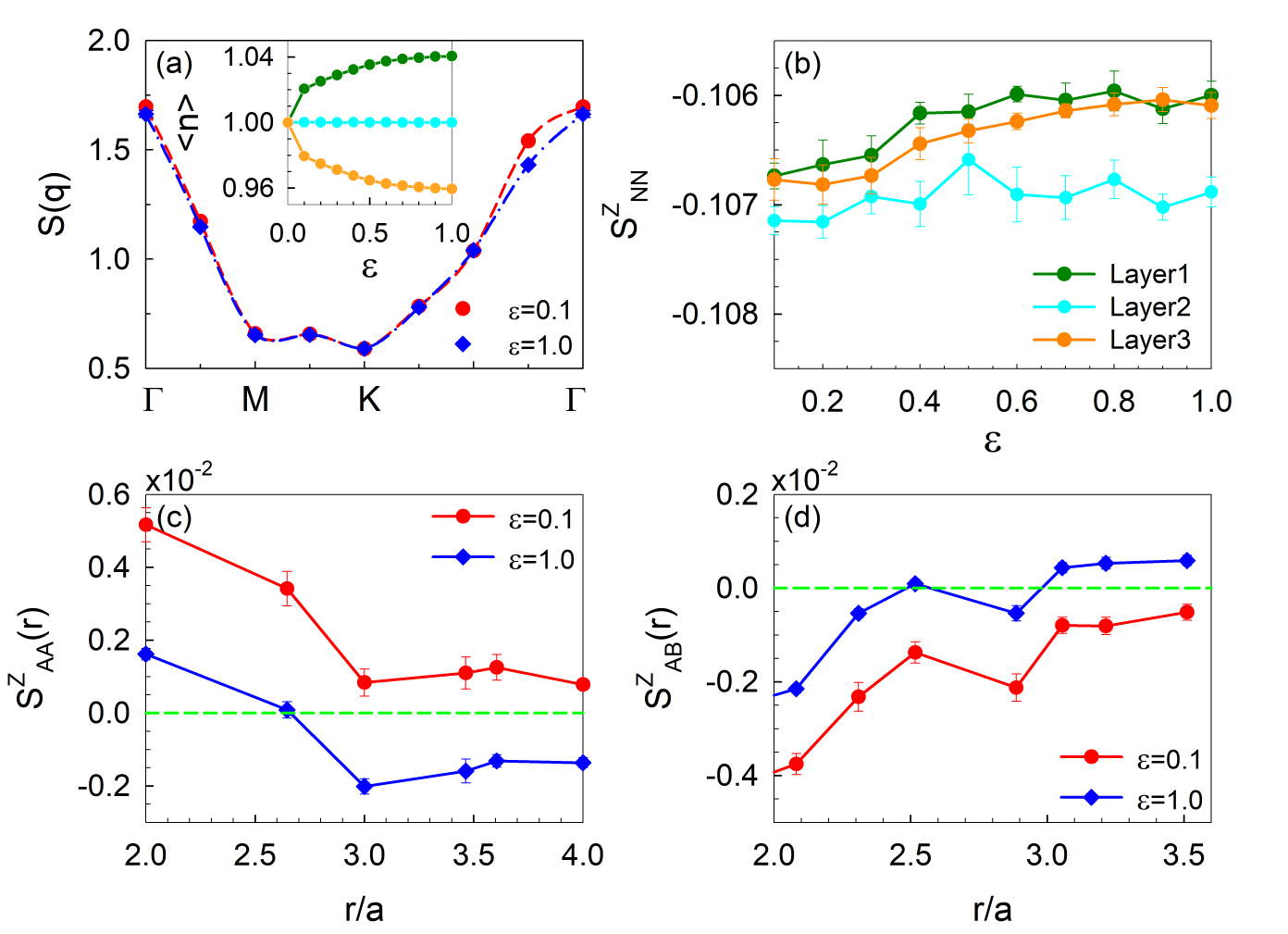}
\caption{(a) Spin structure factor $S(q)$ is shown as a function of momentum at $U = 3$ on the $L = 4$ lattice.
(b) Intralayer NN spin correlation function as a function of potential difference $\varepsilon$ at $U = 3$.
(c) Intrasublattice and (d) intersublattice long-range spin correlations versus distance $r/a$. Results are obtained at $U=3$, $L=4$.
The green dashed lines indicate the position of $0.0$.
The inset: averaged electron density of each layer $\langle n \rangle$ versus potential difference $\varepsilon$. }
\label{Magnetic}
\end{figure}

Next, we discuss the impact of the electric field on the magnetic property of the studied system.
To examine how the magnetic order develops at half filling, we compute the antiferromagnetic (AFM) spin structure factor, which is defined as
\begin{equation}
S(q) = \frac{1}{N_s} \sum_{dd^\prime}\sum_{ij}\sum_{l}{\epsilon}_{dd^\prime}e^{iq(i_{ld}-j_{ld^\prime})}\langle S_{ild}S_{jld^\prime} \rangle,
\label{Sq}
\end{equation}
where $d$ or $d^\prime$ denotes the sublattice index. $S_{ild} = n_{ild\uparrow} - n_{ild\downarrow}$, ${\epsilon}_{dd^\prime} = 1$ for $d = d^\prime$, and ${\epsilon}_{dd^\prime} = -1$ for $d \neq d^\prime$.
The peaks in this quantity are related to the dominant spin ordering. As shown in Fig. \ref{Magnetic}(a), for both $\varepsilon = 0.1$ and
$\varepsilon = 1.0$, we observe a peak at the $\Gamma$ point related to AFM fluctuations. The association of $\Gamma$ with AFM (rather than
FM) is due to the additional factor $\varepsilon_{d,d^{\prime}}$ in the structure factor, which changes sign in opposite sublattices. Moreover,
this peak is weakly suppressed as $\varepsilon$ changes from $0.1$ to $1.0$.

Figure \ref{Magnetic}(b) shows the intralayer NN spin correlation
\begin{equation}
S_{NN}^Z(l) = \frac{1}{3}\sum_{i,j=i+{\delta}_l}\langle (n_{ild\uparrow}-n_{ild\downarrow})(n_{jld^{\prime}\uparrow}-n_{jld^{\prime}\downarrow}) \rangle.
\label{SNNz}
\end{equation}
where the vectors $\delta_\textbf{l} (\textbf{l} = 1 - 3)$ denote the nearest neighbor (NN) intersublattice connections
and the $d$ or $d^{\prime}$ denotes the sublattice index of these nearest neighbor sites and they always have opposite
values.
The negative value of the NN spin correlation indicates that the system has an AFM fluctuation with a perpendicular electric field, consistent with the conclusion of Fig. \ref{Magnetic}(a).
The trend of NN spin correlation in Fig. \ref{Magnetic}(b) suggests that the amplitudes of spin correlations are gradually suppressed with increasing $\varepsilon$, but the suppression is rather weak even at $\varepsilon = 1.0$.
This result demonstrates that AFM fluctuation is
always the dominant magnetic fluctuation, which is consistent with the spin structure factor in Fig. \ref{Magnetic}(a).
Moreover, the first layer and third layer have a similar upward trend, while the second layer only fluctuates.
The layer symmetry is broken by the electric field as observed from the averaged electron density of each layer$\langle n \rangle$ shown in the inset of Fig. \ref{Magnetic}(a).
The doping density of electrons in the first layer is the same as the doping density of holes in the third layer, resulting in their similar behavior, but the second layer remains half filled.

To determine whether the long-range AFM fluctuation survives in the electric field, in Figs. \ref{Magnetic}(c) and \ref{Magnetic}(d), we plot the intrasublattice spin correlations
\begin{equation}
S_{AA}^Z(\textbf{R} = {\textbf{R}}_{\textbf{i}} - {\textbf{R}}_{\textbf{j}}) = \frac{1}{3}\sum_{l} \langle (n_{ilA\uparrow}-n_{ilA\downarrow})(n_{jlA\uparrow}-n_{jlA\downarrow}) \rangle
\label{SAA}
\end{equation}
and intersublattice spin correlations
\begin{equation}
S_{AB}^Z(\textbf{R} = {\textbf{R}}_{\textbf{i}} - {\textbf{R}}_{\textbf{j}}) = \frac{1}{3}\sum_{l} \langle (n_{ilA\uparrow}-n_{ilA\downarrow})(n_{jlB\uparrow}-n_{jlB\downarrow}) \rangle
\label{SAB}
\end{equation}
as a function of distance $r/a$, where a is the lattice constant.
At $\varepsilon = 0.1$, $S_{AA}^Z(r)$ displays positive values and $S_{AB}^Z(r)$ displays negative values, which reveals the existence of long-range AFM ordering.
At $\varepsilon = 1.0$, with the increasing distance $r/a$, $S_{AA}^Z(r)$ decreases from positive to negative and $S_{AB}^Z(r)$ grows from negative to positive,
indicating that the AFM ordering is suppressed as the electric field increases and becomes destroyed when $r/a > 2.5$.

\begin{figure}[tbp]
\includegraphics[scale=0.50]{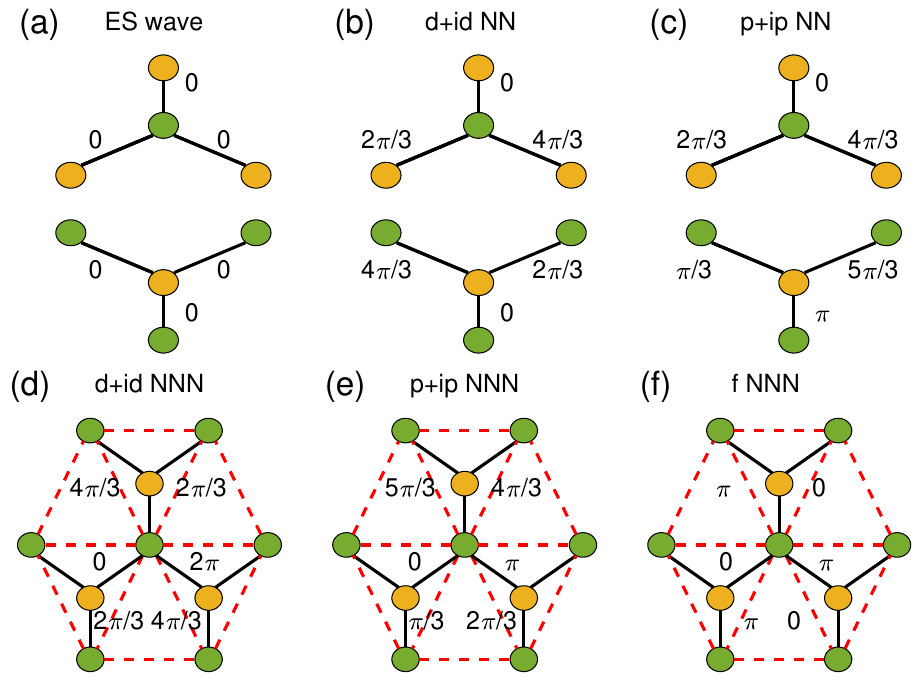}
\caption{Phases of the pairing symmetries on the honeycomb lattice: (a) $ES$ wave, (b) $d + id$ with NN, (c) $p + ip$ with NN, (d)$d + id$ with NNN, (e) $p + ip$ with NNN, and (f) $f$ NNN wave.
Here, the different colored dots denote sites of the different sublattices A and B.
} \label{Symmetries}
\end{figure}

To investigate the superconducting property, we studied the pairing correlations for various pairing symmetries, which are defined as
\begin{equation}
C_\alpha(\textbf{R} = {\textbf{R}}_{\textbf{i}} - {\textbf{R}}_{\textbf{j}}) = \frac{1}{3} \sum_{l} \langle \Delta_{l\alpha}^\dagger(i)\Delta_{l\alpha}(j)\rangle,
\label{Cpaire}
\end{equation}
where $\alpha$ stands for different pairing symmetries.
Due to the constraint of the on-site Hubbard interaction in Eq.(\ref{1}),
the local pairing should be suppressed by the Coulomb repulsion;
we consider the pairing order parameter $\Delta_{l\alpha}^\dagger(i)$ of nearest neighbor bonds and next nearest neighbor bonds and
the nearest neighbor bonds pairing is defined as follows:
\begin{equation}
\Delta_{l\alpha}^{\dagger}(i) = \sum_{\textbf{l}}f_{\alpha}^{\dagger}(\delta_{\textbf{l}})(a_{li\uparrow}b_{li+\delta_{\textbf{l}}\downarrow} \pm a_{li\downarrow}b_{li+\delta_{\textbf{l}}\uparrow})^{\dagger},
\label{Delta}
\end{equation}
The next nearest neighbor bonds pairing is
\begin{equation}
\Delta_{l\alpha}^{\dagger}(i) = \sum_{\textbf{l}}f_{\alpha}^{\dagger}(\delta_{\textbf{l}})(a_{li\uparrow}a_{li+\delta_{\textbf{l}}\downarrow} \pm a_{li\downarrow}a_{li+\delta_{\textbf{l}}\uparrow})^{\dagger},
\label{Delta2}
\end{equation}
where $f_\alpha(\delta_\textbf{l})$ in the pairing function is the form factor distinguishing different pairing symmetries
and the $-$($+$) sign is associated with spin-singlet(triplet) pairing.
Here, in Eq. (\ref{Delta}), the vectors $\delta_\textbf{l} (\textbf{l}=1-3)$ denote the nearest neighbor (NN) intersublattice connections (the sublattice index $m \ne n$)
and, in Eq. (\ref{Delta2}), $\delta_\textbf{l} (\textbf{l} = 1 - 6)$ denotes the next nearest neighbor (NNN) intrasublattice connections (the sublattice index $m = n$), as sketched in Fig. \ref{Symmetries}.

Considering the special structure of the honeycomb lattice, three possible NN pairing symmetries are characterized by (a) extended $S (ES)$, (b) $d + id$ and (c) the $p + ip$ wave\cite{PhysRevB.94.115105,PhysRevB.97.075127,PhysRevB.84.121410,PhysRevB.90.245114,HUANG2019310,Ma2015EPL}.
These extended pairing symmetries are defined with different phase shifts upon $\pi/3$ or $2\pi/3$ rotations.
The singlet $ES$ wave and NN-bond $d + id$ pairing have the following form factors:
\begin{equation}
f_{ES}(\delta_\textbf{l}) = 1, \textbf{l} = 1,2,3,
\label{8}
\end{equation}
\begin{equation}
f_{d+id}(\delta_\textbf{l}) = e^{i(\textbf{l}-1){\frac{2\pi}{3}}}, \textbf{l} = 1,2,3,
\label{9}
\end{equation}
For the NN-bond $f_{p+ip}$ pairings, the form factors of the A and B sublattices are different, where
\begin{equation}
f_{p+ip}(\delta_\textbf{al}) = e^{i(\textbf{l}-1){\frac{2\pi}{3}}}, \textbf{l} = 1,2,3,
\label{10}
\end{equation}
\begin{equation}
f_{p+ip}(\delta_\textbf{bl}) = e^{i[(\textbf{l}-1){\frac{2\pi}{3}}+\pi]}, \textbf{l} = 1,2,3.
\label{11}
\end{equation}
for A and B, respectively, which are similar except that there is a $\pi$ phase shift.
We also considered three common NNN bond pairings: (d) $d + id$, (e) $p + ip$, and (f) $f$ wave symmetry\cite{PhysRevB.97.075127,Ma2015EPL}, which have the following form factors:
\begin{equation}
f_{d+id}(\delta_\textbf{l}) = e^{i(\textbf{l}-1){\frac{2\pi}{3}}}, \textbf{l} = 1,2,3...6.
\label{12}
\end{equation}
\begin{equation}
f_{p+ip}(\delta_\textbf{l}) = e^{i(\textbf{l}-1){\frac{\pi}{3}}}, \textbf{l} = 1,2,3...6.
\label{13}
\end{equation}
\begin{equation}
f_{f}(\delta_\textbf{l}) = e^{i\frac{1+(-1)^{\textbf{l}}}{2}\pi}, \textbf{l} = 1,2,3...6.
\label{14}
\end{equation}

\begin{figure}[tbp]
\includegraphics[scale=0.55]{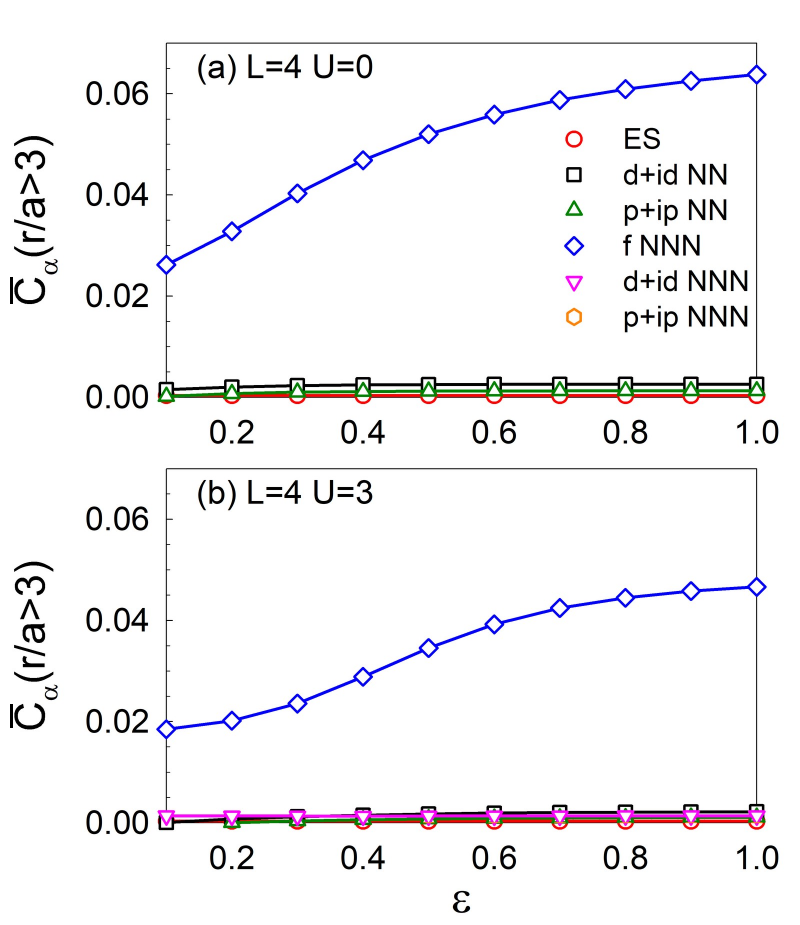}
\caption{Pairing correlations as a function of potential difference $\varepsilon$ at (a) $U = 0$ and (b)$U = 3$ on the $L = 4$ lattice for different pairing symmetries.
} \label{Pair_e}
\end{figure}

Figure \ref{Pair_e} presents the long-range parts of various pairing correlations as a function of $\varepsilon$.
We sum up the correlations whose distance is three times larger than lattice constant $a$
and we denote it by $\bar{C}_\alpha(r/a > 3)$.
The reason that we look at pairing at $r/a > 3$ is that, as $r/a < 3$, in the value of pairing interaction
mix too much contribution from the spin correlation. The spin correlations decrease very fast as the lattice distance
increases. As $r/a > 3$, the contribution from the spin correlation shall have little effect on the dominate pairing
correlation.
One can readily see that the pairing correlation with $f$ pairing symmetry is much larger than other symmetries and it is greatly enhanced with increasing $\varepsilon$.
A comparison between Figs. \ref{Pair_e}(a) and \ref{Pair_e}(b) shows that the $f$ pairing symmetry is slightly reduced as the value of $U$ increases; in addition, the trends for the pairing symmetries are very similar for $U = 0$ and $U = 3$. Besides, one can find that the $\bar{C}_{\alpha}(r/a > 3)$ are enhanced by the increasing electric field. Recalling the suppression of long-range antiferromagnetic fluctuations that have been shown in Figs. \ref{Magnetic}(c) and \ref{Magnetic}(d),
one can conclude that the electric field weakens the long-range antiferromagnetism and enhances the superconductivity.
The competition between superconductivity
and antiferromagnetism is an important topic in strong correlated systems and these results may reveal this
competition in rhombohedral trilayer graphene under an electric field.

\begin{figure}[tbp]
\includegraphics[scale=0.55]{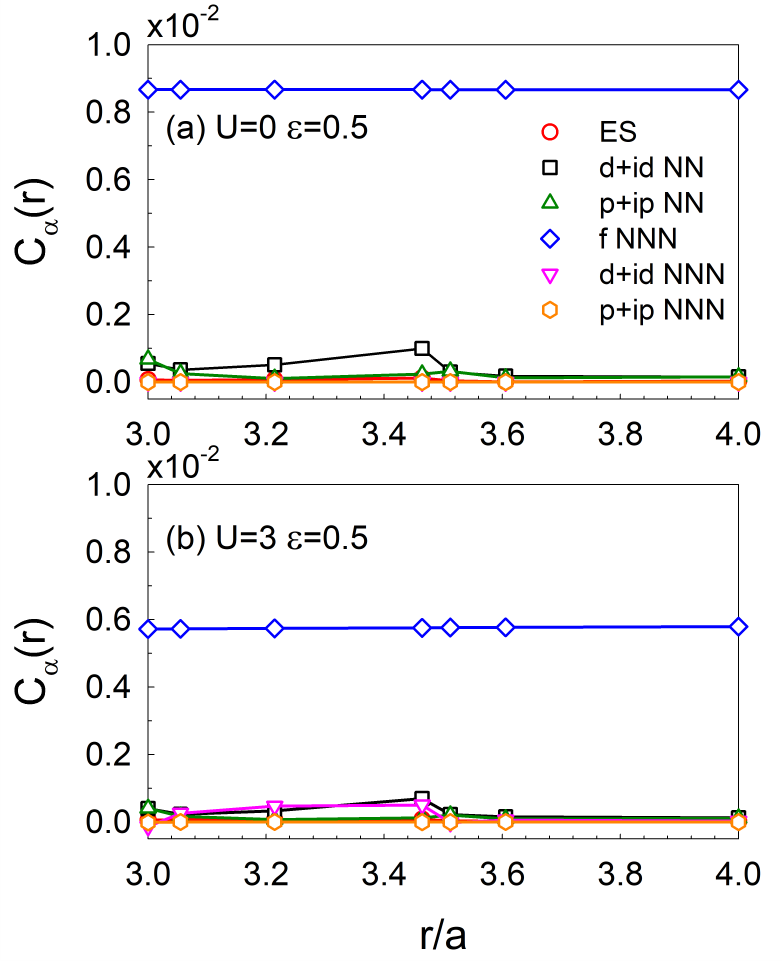}
\caption{Pairing correlations as a function of pairing distance $r/a$ at (a) $U = 0$ and (b)$U = 3$ with $\varepsilon = 0.5$ for different pairing symmetries.
} \label{Pair_r}
\end{figure}

Moreover, Fig. \ref{Pair_r} shows $C_\alpha$ versus distance $r/a$ at $\varepsilon = 0.5$.
For all long-range distances, the superconducting pairing correlation with $f$ wave symmetry is more dominant than that of other pairing symmetries, which confirms the findings in Fig. \ref{Pair_e}.

Based on the above results, our simulation results support the notion  that the system favors $f$ pairing symmetry under the control of a perpendicular electric field.
However, previous quantum Monte Carlo studies on graphene \cite{PhysRevB.97.075127,Fang_2020} have indicated that the conclusion derived from the electron pairing correlations might be misleading due to the noninteracting part of the Hamiltonian.
Comparing Fig. \ref{Pair_e} and Fig. \ref{Pair_r}, one can see that interaction strength $U$ has little impact on the shapes of the pairing symmetries except for slightly suppressing their values, which indicates that $f$ wave symmetry dominates other symmetries, which may be due to the electronic structure of the noninteracting part.

Since the interactions play an important role in the shape of superconductivity,
we are more concerned about the pairing correlations generated from the interactions.
To identify the actual dominant pairing symmetry, we calculated the corresponding vertex contribution, which is defined as follows:
\begin{equation}
V_\alpha(\textbf{R}) = C_\alpha(\textbf{R}) - \widetilde{C}_\alpha(\textbf{R}),
\label{Vpaire}
\end{equation}
where $\widetilde{C}_\alpha(\textbf{R})$ is an uncorrelated single-particle contribution, which is achieved by replacing $\langle a_{li\downarrow}^{\dagger}a_{lj\downarrow}b_{i+\delta_\textbf{l}\uparrow}^{\dagger}b_{j+\delta_{\textbf{l}^{'}}\uparrow}\rangle$ in Eq. (\ref{Cpaire}) with $\langle a_{i\downarrow}^{\dagger}a_{j\downarrow}\rangle \langle b_{i+\delta_\textbf{l}\uparrow}^{\dagger}b_{j+\delta_{\textbf{l}^{'}}\uparrow}\rangle$.
Positive (negative) $V_\alpha(\textbf{R})$ signals an enhanced (suppressed) tendency for the pairing symmetry $\alpha$ and we can determine the dominant pairing from the tendency of the effective pairing correlation function.

\begin{figure}[tbp]
\includegraphics[scale=0.55]{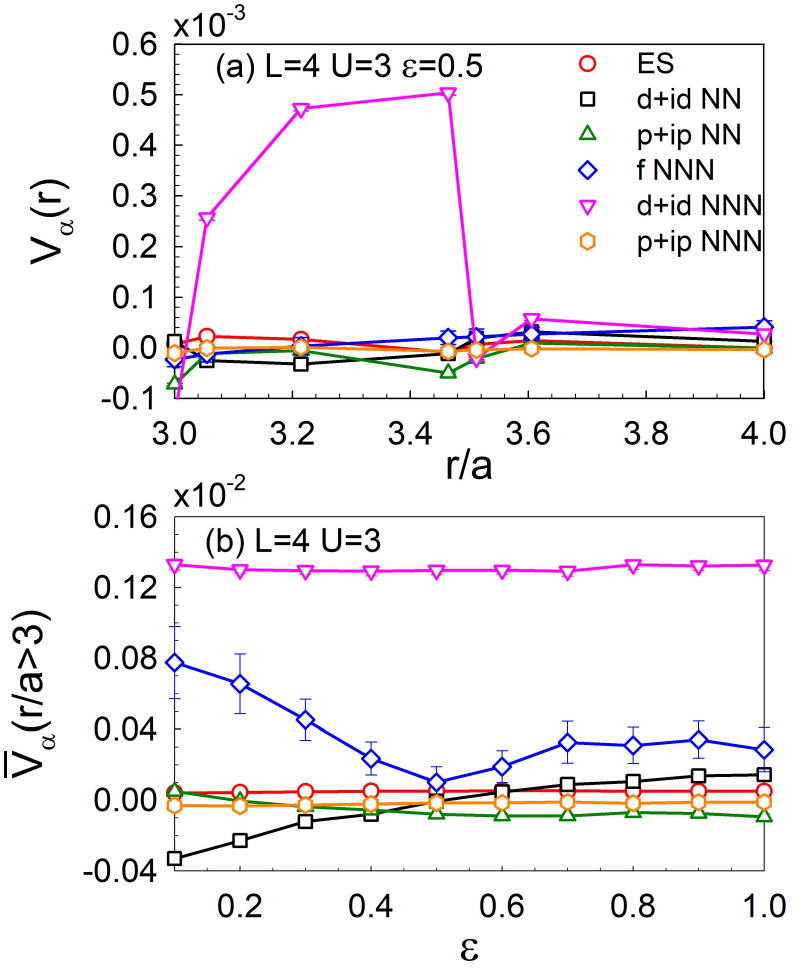}
\caption{Vertex functions as a function of (a) distance $r/a$ and (b) potential difference $\varepsilon$ for different pairing symmetries at $U = 3$.
} \label{Vertex}
\end{figure}

As shown in Fig. \ref{Vertex}(a), the distance-dependent vertex contribution is shown for $\varepsilon = 0.5$.
It is clear that $V_{d+id}(R)$ is larger than the amplitude of other symmetries for all long-range distances between electron pairs, demonstrating that the NNN bond $d + id$ is the dominant pairing symmetry in ABC-TLG under the perpendicular electric field.
The long-range vertex contribution as a function of $\varepsilon$ is also shown in Fig. \ref{Vertex}(b).
One can see that $\varepsilon$ leads the vertex contribution to fluctuate, but the leading pairing symmetry does not change for all $\varepsilon$.
Similar to the correlations, $\bar{V}_\alpha(r/a > 3)$ is defined to demonstrate the long range part of the vertex; we sum up
the vertex with $r/a > 3$, where $a$ is the lattice constant.
The values of $\bar{V}_\alpha(r/a > 3)$ for the $d + id$ wave and $f$ wave remain positive as $\varepsilon$ varies.
The positive effective pairing interaction indicates that there actually exists
attraction for the $d + id$ and $f$ pairing symmetries.
Moreover, the $d + id$ symmetry is almost insensitive to the electric field, while the $f$ symmetry gradually decreases under the influence of the electric field.
The $d + id$ NN wave gradually increases from a negative value to zero with increasing $\varepsilon$ and $\varepsilon$ has a negligible effect on the vertex contribution for other symmetries whose corresponding values are close to zero, suggesting that it is difficult to form an effective pairing attraction for these pairing symmetries.

\begin{figure}[tbp]
\includegraphics[scale=0.55]{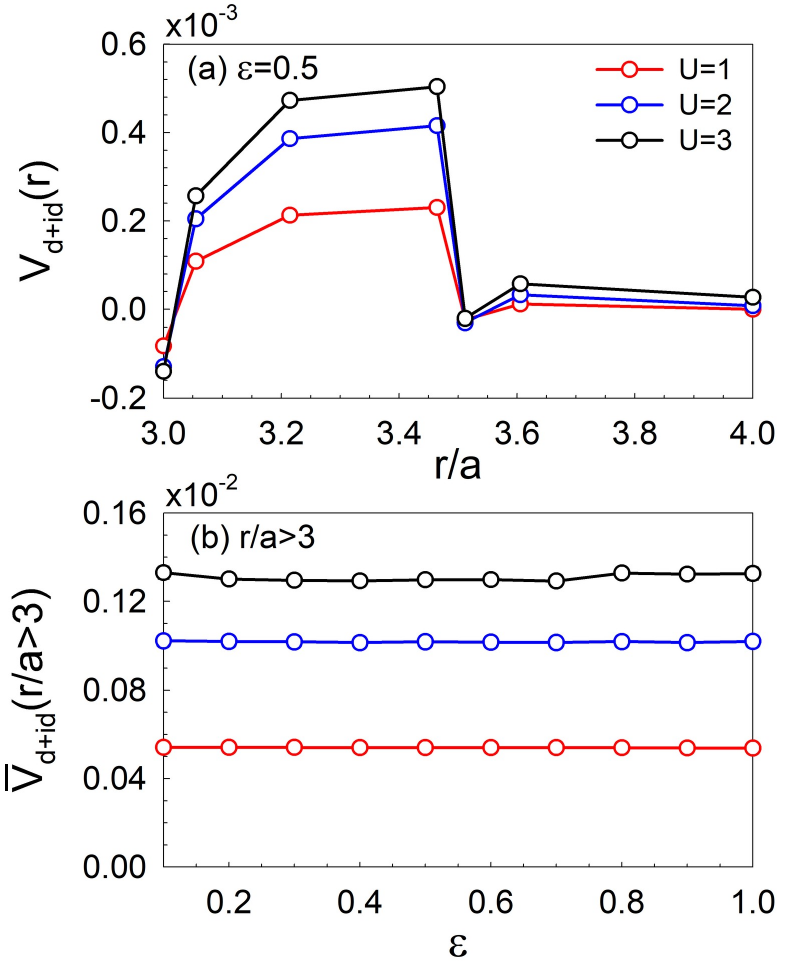}
\caption{Vertex function of $d + id$ pairing symmetry as a function of (a) distance $r/a$ and (b) potential difference $\varepsilon$ for different on-site interaction U.
} \label{Vertex_d}
\end{figure}

To learn more about the NNN-$d + id$ vertex contribution, we examined the evolution of $V_{d+id}$ with various on-site Coulomb interactions and electric fields, as shown in Fig. \ref{Vertex_d}.
We observe that the value of the vertex contribution shows a strong increase as $U$ is added in Fig. \ref{Vertex_d}(a), indicating the importance of electronic correlation in enhancing the $d + id$ superconducting order, which is consistent with a previous study \cite{PhysRevB.104.035104}.
The results shown in Fig. \ref{Vertex_d}(b) confirm that the long-range part of the vertex contribution with $d + id$ pairing symmetry is almost independent of the electric field.

\noindent
{\it Conclusions}.
In summary, we have studied the magnetic properties and pairing symmetry of half filled rhombohedral trilayer graphene under an electric field.
Our simulations based on the Hubbard model indicate that the system exhibits an antiferromagnetic correlation, which is slightly suppressed with increasing electric field.
At half filling, the superconducting pairing with $d+id$ symmetry dominates over other pairing symmetries.
We have also analyzed the effect of the on-site interaction and electric field on the superconductivity.
It is found that the dominant $d+id$ superconducting pairing is enhanced as the on-site interaction increases and is robust against variations in the electric field, which supports the scenario of superconductivity originating from strong electronic correlations.
Our intensive numerical results unveil a possible interaction driven superconductivity with $d + id$ pairing symmetry
in rhombohedral trilayer graphene under an electric field.


This work was supported by the NSFC (Grant No. 11974049). The numerical simulations were performed at the HSCC of Beijing Normal University and used Tianhe-2JK in the Beijing Computational Science Research Center.

\appendix
\noindent
\section{  Appendix}

\begin{figure}[htbp]
\includegraphics[scale=0.35]{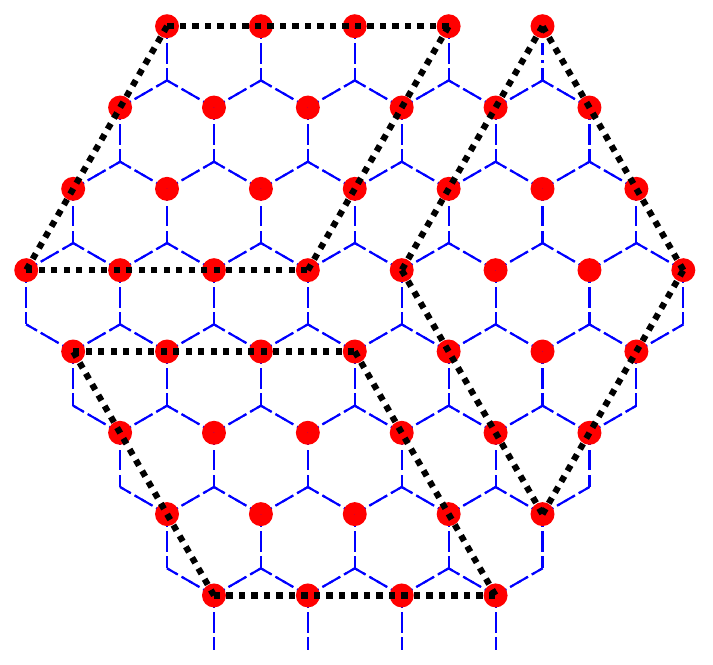}
\caption{Sketch of the triangular sublattice of honeycomb structure for L = 4.
} \label{Unit cell}
\end{figure}

To make the geometry of trilayer graphene more clear than we have built in the main text, we now present the planar structure schematic of the triangular sublattice of honeycomb structure with linear lattice size L = 4 in Fig. \ref{Unit cell}.
The red symbols denote one of the inequivalent triangular sublattices, the underlying honeycomb lattice is represented by the blue dashed lines, and the unit cell forming the triangular lattice is marked by black dotted lines which consist of $L^2$ sites.
Here the designed triangular lattice has $3\times L^2$ sites, which is a $1/2$ subset of the honeycomb lattice, so the total sites of trilayer graphene could be expressed as $N_S = 3\times 2\times 3\times L^2$.

\begin{figure}[htbp]
	\includegraphics[scale=0.35]{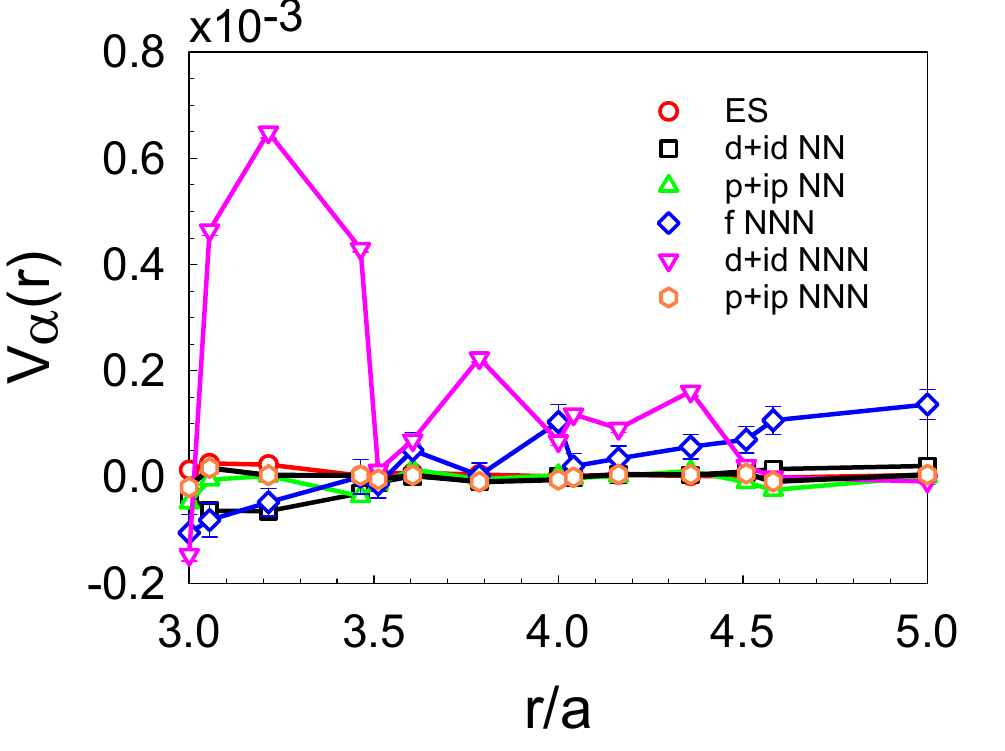}
	\caption{Pairing vertex at $L=5$, $U=3.0$ and $\varepsilon=0.5$, where the results
		are qualitatively the same as $L=4$.}
	\label{l5vertex}
\end{figure}

To check the consistency at different lattice sizes, we perform simulations on $L=5$.
We can see that, in Fig. \ref{l5vertex}, the dominance of $d+id$ remains unchanged.

\begin{figure}[htbp]
\includegraphics[scale=0.35]{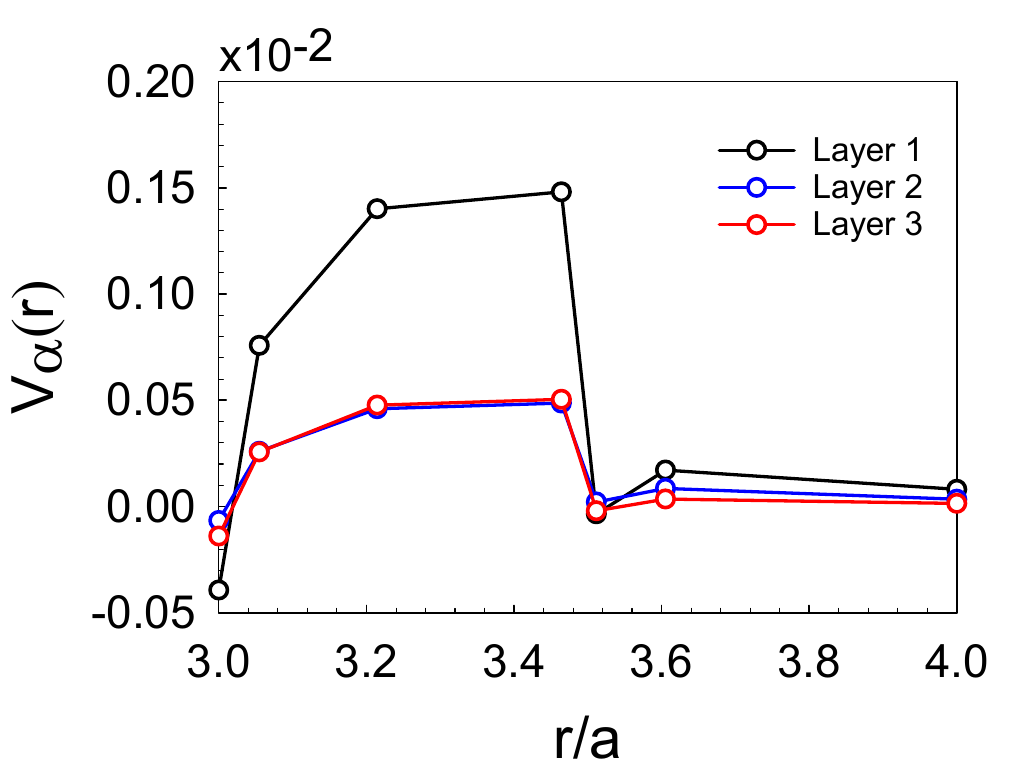}
\caption{Pairing vertex of different layers at $U=3.0$ on the $L=4$ lattice, where we can see that the first layer
is larger than the others. However, the tendencies of them are similar to each other.}
\label{sclayer}
\end{figure}

Finally, we check the layer dependence of pairing correlations in  Fig. \ref{sclayer}; the second and the
third layer are nearly the same and the first layer is larger. However, the tendencies of them
are quite similar, so, in the main text, we average over them for convenience.

\bibliography{ref}
\end{document}